\documentstyle[prl,aps,twocolumn,epsfig]{revtex}
\title{A classical field method for time dependent Bose condensed gases}
\author{ Alice Sinatra, Carlos Lobo$^*$, Yvan Castin  \\ {\small Laboratoire
Kastler Brossel, 24 Rue Lhomond, 75231 Paris Cedex 05, France}\\ {${}^*$ \small
University of Illinois at Urbana-Champaign,  1110 West Green Street, Urbana,
Illinois 61801-3080} }
\begin{document}
\maketitle
\begin{abstract} 
We propose a method to study the time evolution of
Bose condensed gases perturbed from an initial thermal equilibrium,
based on the Wigner representation of the $N$-body density operator.
We show how to generate a collection of random classical fields 
sampling the initial Wigner distribution in the number conserving
Bogoliubov approximation. The fields are then evolved with the
time dependent Gross-Pitaevskii equation. We illustrate the method
with the damping of a collective excitation of a one-dimensional
Bose gas.
\end{abstract}

Since the first experimental demonstrations of Bose-Einstein condensation in
atomic gases \cite{revue}, the role played by finite temperature
effects
in the physics of Bose condensed alkali gases  has drawn increasing
attention. For
example, in thermal equilibrium, both the spatial density of the condensate
atoms
\cite{modif_dens} and the distribution of number of particles in the condensate
are modified due to the presence of the thermal atoms \cite{Scully}.
Similarly, we must take into account the noncondensed atoms in order to
explain
time-dependent phenomena like the damping and frequency shifts of collective
modes \cite{JILAcoll,MITcoll}, or the evolution of the recently created
vortices
\cite{vortex}, where the dissipation of the condensate motion is
provided by the noncondensed atoms \cite{Giora}.

The widely used Gross-Pitaevskii equation,
the nonlinear Schr\"odinger equation for the condensate wave function, 
is not able to
describe
these effects since it neglects the interaction between the condensate and the
noncondensed atoms \cite{Pines}. 
One possibility to go beyond the Gross-Pitaevskii equation
is to use Bogoliubov theory, which is a perturbative
method valid for a small noncondensed fraction.
The Bogoliubov method can be applied to thermal equilibrium but also
to time-dependent situations: 
in a $U(1)$ symmetry breaking point of view, 
to zeroth order one solves the time-dependent
Gross-Pitaevskii equation for the condensate field $\psi_0(\vec{r},t)$,  
to first order
one linearizes the quantum field equations around the classical field
$\psi_0(\vec{r},t)$  
to get the dynamics of noncondensed particles,  
to second order one includes the back-action of 
noncondensed particles on the condensate. 
However, in this way one can predict only small corrections
to the Gross-Pitaevskii equation. Additionally, if the number of
noncondensed  particles increases during the evolution of the system, the
Bogoliubov approach is valid only for short times \cite{Yvan_et_Ralph}.
Another existing approach is the mean field Hartree-Fock-Bogoliubov
approximation.
This approach is known however to present consistency problems 
and is still the object of research \cite{MorganLudoMaxim}.
 
In this paper we propose an alternative  method 
to study the time evolution of
Bose condensed gases perturbed from an initial thermal equilibrium, 
based
on the classical field approximation in the Wigner representation, the
so-called truncated Wigner approximation.
The classical field approximation amounts to evolving a set of initially
randomly distributed atomic fields with the usual Gross-Pitaevskii
equation, with the crucial point that the field is now the whole matter field,
not simply the condensate field.
This approximation has been used already in the Glauber-P
representation to study the formation of the condensate
and it is expected to be correct for modes having 
a large occupation number \cite{Kagan,Burnett}. 
 For such modes this approximation is known however to be most
accurate in the Wigner representation as quantum noise is mimicked
by classical noise in the initial state \cite{Drummond}.   
In addition 
in such representation, the approximation
is valid also for modes with a small occupation number
as long as their evolution is well
described by a quadratic approximation to the Hamiltonian, like in
the Bogoliubov approach.  The classical field approximation is indeed
exact in the Wigner representation for a quadratic Hamiltonian!
The truncated Wigner approximation is however expected to fail in describing
the relaxation of the gas to the correct thermal equilibrium,
and one has to supplement the Gross-Pitaevskii equation by kinetic equations
for the high energy modes \cite{Kagan}.

The main result of this paper is the construction of an algorithm
to approximately sample the Wigner distribution at
thermal equilibrium, using the number conserving Bogoliubov theory
\cite{Girardeau,GardinerBogol,CastinDum}. This was the missing block
for the implementation of the truncated Wigner approach to study the dynamics
of low temperature Bose condensed gases.
We apply our method to the damping
of the breathing mode of a 1D condensate in a harmonic trap, 
which exemplifies
the superiority of the truncated Wigner over the
time-dependent Bogoliubov approach.

We recall briefly the number conserving Bogoliubov approach that we use to
describe the initial thermal equilibrium state of the gas. The total number of
particles of the gas is supposed to be fixed and equal to $N$. The existence of
the macroscopically populated mode $\phi(\vec{r}\,)$ motivates the splitting of
the atomic field operator:
\begin{equation}
\hat{\psi}(\vec{r}\,)=\phi(\vec{r}\,) \hat{a}_{\phi}+
\hat{\psi}_{\perp}(\vec{r}\,)
\end{equation} where $\phi$ is the condensate wavefunction, $\hat{a}_{\phi}$
annihilates  a particle in the mode $\phi$. The general idea is to
eliminate the
annihilation operator for the condensate. Unlike the 
usual approach, which replaces $\hat{a}_{\phi}$ by a fixed complex number, we keep
the operator nature of $\hat{a}_{\phi}$ and write it in terms of the
condensate phase operator $\hat{A}_{\phi}$ and the number of condensate
particles
operator $\hat{N}_0$ as:
\begin{equation}
\hat{a}_{\phi} = \hat{A}_{\phi} \hat{N}_0^{1/2} \,.
\end{equation} The modulus of $\hat{a}_{\phi}$ can be re-expressed as a
function
of the number of noncondensed particles:
\begin{equation}
\hat{N}_0 = \hat{N} - \int d\vec{r}\,
\hat{\psi}_{\perp}^{\dagger}(\vec{r}\,)
\hat{\psi}_{\perp}(\vec{r}\,) \,.
\label{eq:def_N0}
\end{equation}

Since $\hat{A}_{\phi}$ and $\hat{A}_{\phi}^{\dagger}$ are commuting
unitary operators \cite{proviso}
\begin{equation}
\hat{A}_{\phi} \hat{A}_{\phi}^{\dagger} = \hat{A}_{\phi}^{\dagger}
\hat{A}_{\phi} = \mbox{Id} 
\end{equation} commuting also with $\hat{\psi}_{\perp},
\hat{\psi}_{\perp}^{\dagger}$, we realize that the Hamiltonian can be
expressed as a function of the fields $\hat{\Lambda}(\vec{r}\,) =
\hat{A}_{\phi}^{\dagger}
\hat{\psi}_{\perp}(\vec{r}\,)$ and $\hat{\Lambda}^{\dagger}(\vec{r}\,)$.
In the Bogoliubov approximation we can then quadratize the Hamiltonian in terms
of $\hat{\Lambda}$: the terms linear in $\hat{\Lambda}$ vanish when $\phi$
solves
the time independent Gross-Pitaevskii equation:
\begin{equation}
H_{\mbox{\scriptsize gp}} \ \phi \equiv 
\left[-\frac{\hbar^2}{2m} \Delta+U(\vec{r}\,) +Ng|\phi|^2 -\mu
\right]\phi = 0 
\end{equation}
where $m$ is the atomic mass, $U(\vec{r}\,)$ is the trapping potential,
$g$ is the atomic coupling constant, and $\mu$ is the
chemical potential.
The resulting quadratic form is the Bogoliubov Hamiltonian
\begin{equation}
\hat{H}_{\rm quad}(\hat{\Lambda})
=\int d^3\vec{r}\,\frac{1}{2}(\hat{\Lambda}^{\dagger},-\hat{\Lambda})
\cdot \mathcal{L}
\pmatrix{\hat{\Lambda} \cr
\hat{\Lambda}^{\dagger} \cr} \label{eq:Hquad}
\end{equation} where the differential operator ${\cal L}$ is given by:
\begin{equation}
\mathcal{L}=
\pmatrix{H_{\mbox{\scriptsize gp}} +QNg|\phi|^2Q & QNg\phi^2 Q^* \cr
         -Q^*Ng(\phi^*)^2Q & -H^*_{gp}-Q^*Ng|\phi|^2Q^* \cr} \label{eq:L}.
\end{equation}
Here $Q \equiv {\rm Id}-|\phi \rangle \langle \phi |$ projects
orthogonally to $\phi$.

 The initial
$N$-body density operator of the gas at temperature
$T$ is then approximated by
\begin{equation}
\hat{\sigma}(t=0)\simeq \frac{1}{Z} \exp\left[-\frac{1}{k_B T}\,
\hat{H}_{\rm quad}(\hat{\Lambda})\right] \,.
\label{eq:sigma_bog}
\end{equation} The Wigner distribution associated to the $N$-body density
operator
$\hat{\sigma}$ is a functional of a complex classical field
$\psi(\vec{r}\,)$. It
is defined as the Fourier transform of a characteristic function $\chi$:
\begin{eqnarray} W(\psi) &\equiv& \int {\cal D}^2\gamma \;
\chi(\gamma)\, e^{ \int \gamma^* \psi-\gamma\psi^*}
\label{eq:Wigner} \\
\chi(\gamma) &=&
\mbox{Tr}\left[\hat{\sigma} e^{\int \gamma
\hat{\psi}^{\dagger}-\gamma^*\hat{\psi}} \right]
\label{eq:chi}
\end{eqnarray} where $\int{\cal D}^2\gamma$ stands for the functional integral
over the real and the imaginary part of the complex classical field
$\gamma(\vec{r}\,)$ and the integrals in the exponent are over space.

We now proceed with the analytical calculation of the Wigner distribution
of the
density operator (\ref{eq:sigma_bog}). As we did with the atomic field
operator,
we split all the classical fields into components parallel and orthogonal
to the
condensate mode $\phi$, e.g. $\psi(\vec{r}\,)=\psi_{\parallel}\phi(\vec{r}\,)
+\psi_{\perp}(\vec{r}\,)$.

It will prove convenient to use the commutation of
$\hat{a}_{\phi}$ with $\hat{\psi}_\perp$ to rewrite the characteristic
function as
\begin{equation}
\chi(\gamma) =
\frac{1}{2} \left\langle
\left\{ e^{\gamma_{\parallel}
\hat{N}_0^{1/2} \hat{A}_{\phi}^{\dagger} 
-\mbox{\scriptsize h.c.}}
, 
e^{\int \gamma_{\perp}
\hat{\Lambda}^{\dagger}\hat{A}_{\phi}^{\dagger}
-\mbox{\scriptsize h.c.}}
\right\}
\right\rangle
\label{eq:def_chi_utile}
\end{equation} where $\{,\}$ stands for the anticommutator and
the brackets denote the expectation value in the
density operator $\hat{\sigma}$.
We then
introduce an
approximation for $\hat{N}_0$  in (\ref{eq:def_chi_utile}) 
replacing the operator 
$\hat{N}$  in Eq.(\ref{eq:def_N0}) by its actual value $N$ in the gas:
\begin{equation}
\hat{N}_0\simeq N -\int \hat{\Lambda}^{\dagger} \hat{\Lambda}.
\label{eq:approx}
\end{equation} As a consequence the phase operators
$\hat{A}_{\phi},\hat{A}^{\dagger}_{\phi}$  commute with all the operators
appearing in $\chi$. We now calculate the trace in
(\ref{eq:def_chi_utile})  using a simultaneous eigenbasis of the phase operator
$\hat{A}_{\phi}$, with eigenvalue $e^{i\theta}$, and of the Bogoliubov
Hamiltonian whose eigenvalues depend on the occupation numbers of elementary
excitations but {\it not} on the phase $\theta$. The form obtained for $W$
clearly preserves the $U(1)$ symmetry:
\begin{equation} W(\psi) \simeq \int_0^{2\pi} \frac{d\theta}{2\pi} \; W_0(\psi
e^{i\theta}).
\end{equation}
$W_0$ is the $\theta=0$ component of the Wigner distribution. By
integrating over the real and the imaginary parts of
$\gamma_{\parallel}$ and by using the cyclic properties of the trace we get
\begin{equation} W_0(\psi)=\delta(\psi_\parallel^I)\,
\mbox{Wig}_{\Lambda}\left[\frac{1}{2} \left\{
\delta(\psi_\parallel^R-\hat{N}_0^{1/2}),\hat{\sigma} \right\} \right]
(\psi_{\perp}) 
\end{equation} where $\psi_{\parallel}^{R,I}$ are the real and imaginary
parts of
$\psi_\parallel$. The expression
$\mbox{Wig}_{\Lambda}\left[f(\hat{\Lambda})\right]$
denotes the Wigner transform with respect to $\hat{\Lambda}$ of any 
function $f$ of $\hat{\Lambda}$. It is obtained from
(\ref{eq:Wigner}) and (\ref{eq:chi}) by the substitutions:
\begin{equation}
\hat{\sigma}\rightarrow f(\hat{\Lambda}) \ \ \ \ 
\hat{\psi}\rightarrow\hat{\Lambda} \ \ \ \ \psi \rightarrow\psi_{\perp} \ \ \ \
\gamma\rightarrow
\gamma_{\perp}.
\end{equation} It is convenient to rewrite $W_0(\psi)$ as a quasiprobability
distribution for
$N_0\equiv \psi^*_{\parallel} \psi_{\parallel}$ and $\psi_{\perp}$:
\begin{equation} P(N_0,\psi_{\perp})= \mbox{Wig}_{\Lambda}\left[\frac{1}{2}
\left\{
\delta(N_0-\hat{N}_0),\hat{\sigma} \right\} \right] (\psi_{\perp})
\label{eq:wigner_final2} \,.
\end{equation} A first insight into (\ref{eq:wigner_final2}) can be obtained by
looking at its marginals. By integrating over $N_0$:
\begin{equation} P(\psi_{\perp}) \equiv \int d N_0 \;\; P(N_0,\psi_{\perp}) =
\mbox{Wig}_{\Lambda} (\hat{\sigma}) (\psi_{\perp})
\end{equation} 
we obtain the Wigner distribution for noncondensed modes, which 
is a Gaussian \cite{Kuthai}:
\begin{equation} P(\psi_{\perp}) \propto \exp \left\{ -\int d^3\vec{r}\,
(\psi_\perp^*,\psi_\perp)
\cdot M \pmatrix{\psi_\perp \cr
\psi_\perp^* \cr} \right\}
\label{eq:Ppsi_perp}
\end{equation}
\begin{equation} M \equiv 
\left( \begin{tabular}{rr}
1  & 0 \\
0 & $-$1
\end{tabular}\right)
 \tanh \frac{\mathcal{L}}{2 k_B T}
\end{equation}
where ${\cal L}$ is the Bogoliubov operator (\ref{eq:L}).
The other marginal is obtained by integrating (\ref{eq:wigner_final2}) over $\psi_\perp$:
\begin{equation} P(N_0) \equiv
\int{\cal D}^2\psi_\perp \, P(N_0,\psi_\perp)=
\mbox{Tr}[\delta(N_0-\hat{N}_0)
\,\hat{\sigma}].
\label{eq:PN_0}
\end{equation}
$P(N_0)$ is then simply the probability distribution of the number of particles
in the condensate \cite{note}.

Let us now discuss the limits of validity of (\ref{eq:wigner_final2}). 
The spectrum of $\hat{N}_0$ contains integer numbers only so
that $\delta(N_0-\hat{N}_0)$ in Eq.(\ref{eq:wigner_final2}) vanishes
when $N_0$ is not integer.
As a function of $N_0$,
$P(N_0,\psi_{\perp})$ is therefore
a ``comb" of delta functions
centered on integer values. The exact Wigner distribution
should be
instead a smooth function of its variables. The comb-artifact comes from the
approximation (\ref{eq:approx}) that we have made on the condensate mode. Our
description is sensible if there are many peaks within the width of the
distribution of $N_0$ for a given $\psi_{\perp}$. 
The variance of $N_0$ for a given $\psi_\perp$, 
that is the conditional variance of $N_0$,  should then satisfy:
\begin{equation}
\mbox{Var}(N_0|\psi_{\perp}) \gg 1 \,.
\label{eq:validity}
\end{equation} We estimate this variance for a ``typical" $\psi_{\perp}$
by taking the mean of $\mbox{Var}(N_0|\psi_{\perp})$
over the probability
distribution of $\psi_\perp$ given in (\ref{eq:Ppsi_perp}):
\begin{equation}
\langle \mbox{Var}(N_0|\psi_{\perp}) \rangle_{\psi_{\perp}} =
\frac{1}{8} \,  \mathrm{Tr}_{\perp} \left[ \mbox{Id} - M^2 \right] \,
\label{eq:validity2}
\end{equation} where the trace is over the Bogoliubov modes.
For the modes with energy much larger than $k_B T$
one has
$M^2 \simeq \mbox{Id}$, while for the modes with energy lower or equal to
$k_B T$
one has $M^2 \ll \mbox{Id}$, so that:
\begin{equation}
\langle \mbox{Var}(N_0|\psi_{\perp}) \rangle_{\psi_{\perp}} \simeq \frac{1}{4}
{\cal N}_{\mbox{\scriptsize th}}
\end{equation} where ${\cal N}_{\mbox{\scriptsize th}}$ is the number of
thermally
populated modes which we require to be large. In practice $k_B T$ should be high enough: for a condensate in an isotropic harmonic
trap of frequency
$\omega$, we should have $k_B T \gg \hbar \omega$.

The next point is to sample the Wigner function (\ref{eq:wigner_final2}). We
proceed in two steps: (i) First we sample  $\psi_{\perp}$ irrespectively to the
value
$N_0$ according to the Gaussian distribution (\ref{eq:Ppsi_perp}) with a
stochastic field method \cite{Kuthai,nouveau}.
(ii) We have to sample the conditional distribution of $N_0$
for a
fixed
$\psi_{\perp}$:
\begin{equation} P(N_0|\psi_\perp) = P(N_0,\psi_\perp) / P(\psi_{\perp}) \,.
\label{eq:PN_0cond}
\end{equation}
It turns out that the width in $N_0$
of the distribution $P(N_0|\psi_\perp)$ is much
narrower than the width of $P(N_0)$ when the condition (\ref{eq:validity}) is
satisfied. In fact as soon as a sufficient number of modes is populated, the
following inequality holds involving the number of modes
${\cal N}_{\mbox{\scriptsize th}}$ with energy lower or equal to $k_BT$, the
average of the number of particles out of the condensate $\langle N - \hat{N}_0
\rangle$, and its variance $\mbox{Var}(N - \hat{N}_0)$:
\begin{equation} {\cal N}_{\mbox{\scriptsize th}} \le  \langle N - \hat{N}_0
\rangle \ll
\mbox{Var}(N - \hat {N}_0) = \mbox{Var}(N_0) \,.
\end{equation} We then replace (\ref{eq:PN_0cond}) with a delta distribution
centered on the mean value of $N_0$ for a fixed $\psi_{\perp}$:
\[
\mbox{Mean}( N_0| \psi_{\perp})= C - \frac{1}{2} \int (\psi_\perp^*,\psi_\perp)
\left[\mbox{Id}-M^2 \right] \pmatrix{\psi_\perp \cr
\psi_\perp^* \cr}
\] where $C$ is a constant \cite{P(N_0)}.

Finally, for each randomly generated $\psi_{\perp}$ and $N_0$ we are 
now able to form the total field $\psi$ as:
\begin{equation}
\psi=N_0^{1/2} \left(\phi + \frac{\phi^{(2)}}{N} \right) + \psi_{\perp} \,.
\label{eq:psi_tot}
\end{equation} 
In (\ref{eq:psi_tot}) the function $\phi^{(2)}/N$ is a correction to
the condensate wavefunction induced by the back-action of noncondensed
particles onto the condensate. This correction has to be included 
since it gives a 
contribution to observables on the same order as $\psi_{\perp}$ 
\cite{CastinDum}. We calculate $\phi^{(2)}/N$ using the procedure of
\cite{Kuthai}. 

In the classical field approximation 
each field $\psi$
evolves according to the time-dependent Gross-Pitaevskii equation
\cite{Drummond,Kuthai}:
\begin{equation} i \hbar \; \partial_t \psi = \left[-\frac{\hbar^2}{2m}
\Delta+U(\vec{r}\,,t) +g|\psi|^2 \right] \psi.
\label{eq:tdgpe}
\end{equation} 
The average of an observable at time $t$ will be obtained by
averaging over many stochastic classical fields $\psi(t)$.

\begin{figure}[htb]
\hskip -11mm
\vspace*{-3mm}
\epsfig{file=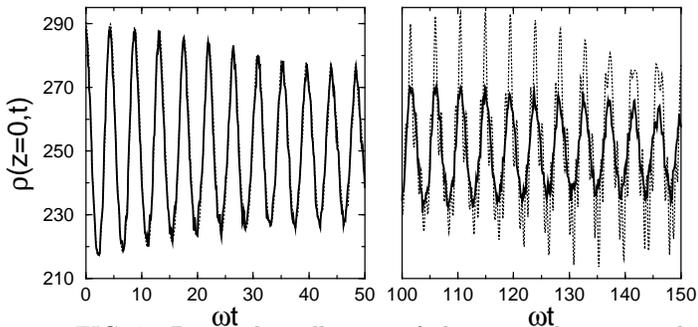,height=4.5cm}
\caption{Damped oscillations of the mean density 
in the center of a 1D condensed cloud
in a harmonic trap after an abrupt change of the trap frequency
$\omega\rightarrow 0.8\omega$. $k_B T=30 \hbar \omega$,
$\mu=3.1 \hbar \omega$ and $N=10\protect^3$. Solid line: Wigner simulation.
Dotted curve: Bogoliubov theory. $\rho$ is in units of $(m\omega/\hbar)^{1/2}$.
}
\label{fig:oscill}
\end{figure}

We now give an illustration of our method.
In Figure~\ref{fig:oscill} we show density oscillations in the center of a
1D Bose condensed gas induced by an abrupt change of the 
harmonic trap frequency $\omega \rightarrow 0.8 \omega$. 
While the Gross-Pitaevskii equation for
the condensate field predicts undamped oscillations,
we get damping at finite temperature when we {\sl average} over
stochastic realizations in the Wigner method \cite{collapse}. 
The time-dependent Bogoliubov theory in \cite{CastinDum}
correctly predicts the damping at short times but gives unphysically
large oscillations at longer times.
We understand this failure as due to the growth in time 
of the number of noncondensed particles and of the condensate wavefunction
correction $\phi^{(2)}$.  This growth
invalidates the linearized treatment at the basis of the time-dependent 
Bogoliubov approach \cite{Yvan_et_Ralph},
at a relatively short time here because of the small number
of particles.
On the contrary, in the truncated Wigner approach, no linearized treatment
is performed and the whole field 
$\psi$ is evolved according to the fully non-linear equation (\ref{eq:tdgpe}).
In Figure 1 a spatial grid of 64 point has been used for discretization in 
the numerical implementation of our algorithm. 
We have tested that the results are unchanged by doubling the number 
of points in the grid \cite{cutoff}. 

In conclusion we have developed an efficient 
algorithm for approximate stochastic sampling of the Wigner
representation of the $N$-body density matrix of a Bose condensed
gas in thermal equilibrium. 
The initial distribution of atomic fields $\psi$ given by this sampling
can then be evolved in the classical
field (truncated Wigner) approximation to study the response
of the gas to an excitation.
As far as time-independent situations are concerned, our approach is
equivalent to the number conserving  Bogoliubov approach.
In the time dependent case however
unlike the Bogoliubov approach, the  
classical field evolution is fully nonlinear both for the condensate and the
noncondensed particles, which extends the applicability of the method 
to longer evolution times.

We acknowledge useful discussions with Anthony Leggett, Yuri Kagan,
Gora Shlyapnikov and Crispin Gardiner. LKB is a unit\'e de
recherche de l'ENS et de l'Universit\'e Paris 6, associ\'ee au CNRS. This work
was partially supported by National Computational Science 
Alliance under DMR 99 00 16 N. C.L.\ acknowledges support from a 
PRAXIS XXI fellowship.


\begin{thebibliography}{99}
\bibitem{proviso} We neglect here border effects appearing when the condensate
mode is empty.
\bibitem{note} Note that this should not be the case for the exact Wigner distribution:
e.g. $\langle N_0 \rangle$ should be equal to $\langle \hat{N}_0 \rangle +
1/2$;
the variances should differ by $1/4$.
\bibitem{P(N_0)}
This algorithm generates the probability 
distribution of the number of condensed particles $P(N_0)$.
We checked in 1D and 2D that
the first two moments of the obtained 
distributions reproduce the results of the Bogoliubov theory,
which validates the approximation of $P(N_0|\psi_\perp)$
by a $\delta$.
For the ideal Bose gas we successfully tested our result 
for $P(N_0)$ against the exact one, see \cite{nouveau}.
\bibitem{collapse} This damping does not occur for a {\sl single} 
realization: our 1D model exhibits a collapse more than a damping.
\bibitem{cutoff} We believe that the cut-off imposed by discretization 
does not affect the predictions of the truncated Wigner method if:
(i) the spatial step of the grid is smaller than both the healing length $\xi$
and the thermal wavelength $\lambda$, and
(ii) the mean field due to the added noise in the Wigner representation
remains small as compared to the real mean field, 
that is the number of \lq empty' modes in the simulation is smaller than the
number of particles. In 3D all these conditions are compatible for a 
dilute ($\rho |a|^3\ll 1$) and degenerate ($\rho \lambda^3\gg 1$) Bose
gas, where $a$ is the $s$-wave scattering length and $\rho$ the spatial density.
A more detailed discussion is given in \cite{nouveau}.
\end{thebibliography}
\end{document}